**Diffraction characteristics of optical elements designed as phase layers with cosine-profiled periodicity in azimuthal direction**

Suzana Topuzoski[*] and Ljiljana Janicijevic

Institute of physics, Faculty of natural sciences and mathematics, University "Ss. Cyril and Methodius," 1000 Skopje, Republic of Macedonia

*Corresponding author: suzana_topuzoski@yahoo.com

The article concerns an investigation of the Fresnel diffraction characteristics of two types of phase optical elements, under Gaussian laser beam illumination. Both elements provide an azimuthal periodicity of the phase retardation. The first element possess azimuthal cosine-profiled phase changes deposited on a plane base. The second element is a combination of the first element and a thin phase axicon.

The cosine profile of the phase retardation, of both diffractive elements, produces an azimuthal cosine-profiled modulation on their diffractograms. It destroys the vortex characteristics of their diffraction fields.

## 1. INTRODUCTION

A circular Ronchi plate and spiral grating, accurately constructed, can produce an image pattern, centred on the line which is observed over a considerable distance [1]. Durnin showed that optical fields with Bessel function transverse amplitude profile can exhibit nondiverging (or "nondiffracting") properties during propagation, i.e., can preserve the form and the dimensions of their transverse intensity profile [2, 3]. Mathematically, the Bessel beam carries an infinite power over an infinite area, but, its finite-aperture approximations can possess nondiverging characteristics over long distances, as was shown by numerical simulations [2] and analytically [4] with the assumption of a Gaussian aperture. The non-spreading property of the central bright or dark core around the optical axis, can be used in a number of technical applications: alignment in some optical and mechanical engineering, optical manipulation of small particles, as well as atom guiding [5]. Besides optical elements with circular symmetry having radial variation: circular diffraction gratings, conical lenses or axicons [6, 7], holographic optical elements with transmission function of axicon type [8], and certain zone plates [9], for precise alignment purposes, can also serve phase diffraction gratings with azimuthal symmetrical variation. Thus, the authors in [10] proposed an annular screen in a form of a binary phase daisy with an even number of slices of equal central angle and a rectangular phase profile in azimuthal direction, repeatedly varying between 0 and $\pi$ for the neighbouring pairs of slices. The diffraction patterns in the focal plane and out-of-focus images were recorded experimentally, showing that the irradiance along the optical axis remains equal to zero in both cases.

Diffraction properties of optical fields generated by diffractive, phase-only optical element, whose transmittance is given by a superposition of two spiral phase plates [11, 12] transmission functions: $T(\varphi) = \exp(in\varphi) + \exp(im\varphi)$, were studied in [13]. In the previous equation with $\varphi$ the azimuthal coordinate is denoted, while $n$ and $m$ are integer topological charges of the spiral phase plates. As a result, camomile-shaped diffraction patterns were predicted and observed, with number of petals being equal to $|n - m|$.

In Section 2 of this article we investigate a phase layer with cosine profile, azimuthally varying with a period of one complete azimuthal change being equal to $2\pi/p$ ($p$ is an integer). Its transmission function is found to be equivalent to a composition of a planparalel plate, and a sum of spiral phase layers having topological charges $+mp$ and $-mp$ ($m=1,2,3,...$). Theoretically we study the Gaussian laser beam transformation due to the diffraction by this type of optical element. It looks similarly to the binary Siemens star [14], known after its applications for testing the resolution of printers and optical systems, as well as for evaluating the basic imaging properties of some phase contrast imaging methods in scanning X-ray microscopy [15]. The optical element studied here provides an azimuthal cosine-profiled phase retardation and therefore we refer to it as a cosine-profiled phase Siemens star (CPSS)-Fig. 1. a. A phase mask-Siemens star with continuous sinusoidal azimuthal profile was fabricated by KrF laser irradiation in order to investigate the spatial resolution of the laser induced backside wet etching process, proposed by the authors in [16].

Further, in Section 3 we study a hybrid optical element consisted of a cosine-profiled phase Siemens star and a thin axicon with a base radius $R_0$, which we name as cosine-profiled phase Siemens star axicon (CPSSA). As it will be seen latter there exists a similarity between our results and those of the authors in [17], who proposed a composite hologram with transmittance equal to a linear combination of transmission functions of two helical axicons with opposite topological charges:

$T(r,\varphi) = \exp[i(n\theta+\alpha)]\exp(i2\pi r/r_o) + \exp[-i(n\theta-\alpha)]\exp(i2\pi r/r_o)$. The helical axicon is a hybrid of an axicon and a phase spiral plate [18]. In the previous equation, $n$ is the topological charge value of the helical axicons, $r_0$ is a constant (connected to the axicon parameters), $\theta$ is azimuthal coordinate. This transmission function was written, as a binary phase, to the magneto-optical spatial light modulator, which was then illuminated with a collimated laser beam. The experimentally obtained diffraction pattern looked like a "nondiffracting" circular array of $2n$ bright spots, surrounding the central dark core, which rotates along the $z$-axis (the angle of rotation is changed by varying the angle $\alpha$).

In this work a theoretical analysis of Fresnel diffraction of a Gaussian laser beam by hybrid optical element, which consists of an axicon and a cosine-profiled phase layer, is performed. The wave and the intensity distributions are derived as a sum of nondiverging, coaxial Bessel beams. The diffraction patterns, calculated numerically, show a central dark core surrounded by $2p$ bright spots in a ring array, when the zeroth-diffraction-order beam is eliminated (for a special value of the relief depth or the refractive index of the cosine-profiled phase layer). Otherwise, because the zeroth-diffraction-order beam is described by a Bessel function of zeroth order, when being present, a central bright spot in the diffractogram appears. The propagation intervals and the radii of the nondiverging ring array of spots, as well as of the central bright spot, are also analyzed.

## 2. COSINE-PROFILED PHASE SIEMENS STAR UNDER GAUSSIAN LASER BEAM ILLUMINATION

The cosine-profiled phase Siemens star is a transparent phase layer with an azimuthal periodicity of the thickness or the refractive index (Fig. 1. a). For the case of variable thickness, the depth of the layer relief is changing as

$$d = \frac{c}{2} + \frac{c}{2}\cos(p\varphi), \tag{1}$$

where $c$ is the total depth of the relief, $p$ is an integer showing the number of periods of the phase change (or, spatial frequency) when the azimuthal variable makes one cycle ($0 < \varphi < 2\pi$)-Fig. 1. b. According to that, the period of one complete azimuthal change is $\tau = 2\pi/p$.

The phase retardation introduced by this layer is

$$T(\varphi) = \exp[-ik(n-1)d] = \exp(-ik\delta)\exp[-ik\delta\cos(p\varphi)], \tag{2}$$

where $\delta = \frac{c}{2}(n-1)$, $n$ is the phase layer refractive index for the incident beam of wavelength $\lambda$, while $k = 2\pi/\lambda$ is the wave number.

Applying the Jacoby-Anger identity for the Bessel function [19], the expression (2) can, also, be written as

$$T(\varphi) = \exp(-ik\delta)\left\{J_0(k\delta) + \sum_{m=1}^{\infty}(-i)^m J_m(k\delta)[\exp(+imp\varphi) + \exp(-imp\varphi)]\right\}, \tag{3}$$

or, as: $T(\varphi) = t_0 + \sum_{m=1}^{\infty} t_{+m}\exp(+imp\varphi) + \sum_{m=1}^{\infty} t_{-m}\exp(-imp\varphi)$, with transmission coefficients denoted as

$$t_0 = \exp(-ik\delta)J_0(k\delta), \quad t_{+m} = t_{-m} = (-i)^m \exp(-ik\delta)J_m(k\delta). \tag{4}$$

Thus, the transmission function of this diffractive optical element is equivalent to a composition of a planparalel plate with transmission coefficient $J_0(k\delta)$ and phase retardation $\exp(-ik\delta)$, and a sum of spiral phase layers having topological charges $+mp$ and $-mp$ ($p$ is an integer, $m$=1,2,3,..).

*Diffraction of a Gaussian laser beam by CPSS*

Normally to the plane $\Delta(r,\varphi)$ where the CPSS is situated, and having its waist in that plane, a Gaussian beam $U^i(r,\varphi,0) = \exp(-r^2/w_0^2)$, with $w_0$ being the beam waist radius, is incident. Its propagation axis is passing through the center of this optical element. The diffracted wave field is calculated in the plane $\Pi(\rho,\theta)$, a distance $z$ from the diffractive optical element plane, using the Fresnel-Kirchhoff diffraction integral [20]

$$U(\rho,\theta,z) = \frac{ik}{2\pi z}\exp\left[-ik\left(z+\frac{\rho^2}{2z}\right)\right]\iint_\Delta T(\varphi)U^{(i)}(r,\varphi,0)\exp\left[-i\frac{k}{2}\left(\frac{r^2}{z}-\frac{2r\rho\cos(\varphi-\theta)}{z}\right)\right]r\,dr\,d\varphi,$$

where $\Delta$ is the area of the diffractive optical element contributing to the diffraction.

After involving the expression for the incident Gaussian wave $U^i(r,\varphi)$ and the transmission function of the CPSS (Eq. 3) into the previous integral, the output diffracted wave field can be written as a sum of coaxial zeroth and higher diffraction orders (positive and negative)

$$U(\rho,\theta,z) = U_0(\rho,\theta,z) + \sum_{m=1}^{\infty}[U_{+m}(\rho,\theta,z) + U_{-m}(\rho,\theta,z)], \qquad (5)$$

where we have denoted the zeroth and the higher-diffraction-order fields, respectively, as

$$U_0(\rho,\theta,z) = \frac{ikt_0}{2\pi z}\exp\left[-ik\left(z+\frac{\rho^2}{2z}\right)\right]\int_0^\infty\left\{\exp\left[-\frac{ikr^2}{2}\left(\frac{1}{z}-\frac{2i}{kw_0^2}\right)\right]\int_0^{2\pi}\exp\left[\frac{ik}{z}r\rho\cos(\varphi-\theta)\right]d\varphi\right\}r\,dr,$$

$$U_{\pm m}(\rho,\theta,z) = \frac{ikt_m}{2\pi z}\exp\left[-ik\left(z+\frac{\rho^2}{2z}\right)\right]$$
$$\times \int_0^\infty\left\{\exp\left[-\frac{ikr^2}{2}\left(\frac{1}{z}-\frac{2i}{kw_0^2}\right)\right]\int_0^{2\pi}\exp\left[\frac{ik}{z}r\rho\cos(\varphi-\theta)\right]\exp(\pm im p\varphi)d\varphi\right\}r\,dr.$$

The integration over the azimuthal variable is done through the use of the identity [19]

$$\exp[(ik/z)r\rho\cos(\varphi-\theta)] = J_0(kr\rho/z) + \sum_{s=1}^{\infty}i^s J_s(kr\rho/z)\{\exp[+is(\varphi-\theta)] + \exp[-is(\varphi-\theta)]\}.$$

Whereas, the integration over the radial variable $r$ is performed by using the known Bessel functions integrals [21]

$$\int_0^\infty \exp(-a_0^2 t^2)t^{\nu+1}J_\nu(b_0 t)dt = \frac{b_0^\nu}{(2a_0^2)^{\nu+1}}\exp\left(-\frac{b_0^2}{4a_0^2}\right) \text{ (for } \operatorname{Re}(\nu) > -1; \operatorname{Re}(a_0^2) > 0\text{)}, \qquad (6)$$

and

$$\int_0^\infty \exp(-a_0^2 t^2) t^{\mu-1} J_\nu(b_0 t) dt = \frac{b_0 \sqrt{\pi}}{8 a_0^{3/2}} \exp\left(-\frac{b_0^2}{8 a_0^2}\right) \left[ I_{(\nu-1)/2}\left(\frac{b_0^2}{8 a_0^2}\right) - I_{(\nu+1)/2}\left(\frac{b_0^2}{8 a_0^2}\right) \right] \text{ (for } \mathrm{Re}(\mu+\nu) > 0;$$

$$\mathrm{Re}(a_0^2) > 0), \tag{7}$$

respectively for the zeroth and the higher-diffraction-order wave fields.

In our case: $a_0^2 = ikq(z)/2zq(0) = 1/w_0^2 + ik/2z$, $b_0 = k\rho/z$, and $\mu=2$. In the first integral $\nu=0$, while in the second integral we replace $\nu = \pm mp$. With $q(z)$ the beam complex parameter is denoted.

The solution for the zeroth-diffraction-order wave amplitude is found in the form

$$U_0(\rho, \theta, z) = J_0(k\delta) \exp(-ik\delta) \frac{w_0}{w(z)} \exp\left\{-i\left[k\left(z + \frac{\rho^2}{2R(z)}\right) - \arctan(z/z_0)\right]\right\} \exp\left(-\frac{\rho^2}{w^2(z)}\right), \tag{8}$$

where $w(z) = w_0 \left[1 + (2z/kw_0^2)^2\right]^{1/2} = w_0 \left[1 + (z/z_0)^2\right]^{1/2}$ is the beam transverse amplitude profile radius for the fundamental mode at distance $z$ from its beam waist, $z_0 = kw_0^2/2$ is the Rayleigh distance, $R(z) = z\left[1 + (kw_0^2/2z)^2\right] = z\left[1 + (z_0/z)^2\right]$ is the beam real on-axial radius of curvature.

The final expressions for the diffracted wave field amplitudes of the positive and negative diffraction orders are the following

$$U_{\pm m}(\rho, \theta, z) = (-i)^m J_m(k\delta) \exp(-ik\delta) \frac{\sqrt{\pi}}{2} \frac{w_0}{w(z)} \left\{\frac{1}{w^2(z)} - \frac{ik}{2}\left(\frac{1}{z} - \frac{1}{R(z)}\right)\right\}^{1/2} \exp[-i\phi(z)]$$

$$\times \exp[\pm imp(\theta + \pi/2)] \rho \exp\left(\frac{-\rho^2}{2w^2(z)}\right) \left\{I_{(mp-1)/2}\left(\frac{x}{2}\right) - I_{(mp+1)/2}\left(\frac{x}{2}\right)\right\}. \tag{9}$$

In Eq. (9) we have denoted the argument of the modified Bessel functions as

$$x = \left[\frac{1}{w^2(z)} - i\frac{k}{2}\left(\frac{1}{z} - \frac{1}{R(z)}\right)\right] \rho^2, \tag{10}$$

and the longitudinal phase as: $\phi(z) = k\left\{z + \frac{k\rho^2}{4}\left[\frac{1}{z} + \frac{1}{R(z)}\right]\right\} - \arctan(z/z_0)$.

The zeroth-diffraction-order beam (8) is a basic Gaussian mode, chargeless, whereas, the higher-diffraction-order beams (9) are vortex ones. Since the bright axis of the zeroth-diffraction-order beam disturbs the registration of the vortex cores of the higher diffraction orders, it could be eliminated by choosing the depth of the relief $c$ with such a value that satisfies the following equation

$$k\delta = y_{0,j}, \quad \text{or} \quad c = \frac{2y_{0,j}}{k(n-1)}, \tag{11}$$

where $y_{0,j}$ is the $j$-th zero of the Bessel function $J_0(k\delta)$.

The representation of the higher diffraction orders as vortex beams with phase singularities with topological charges $+mp$ and $-mp$, and through the modified Bessel functions of orders $(mp-1)/2$ and $(mp+1)/2$, is similar to that one concerning the higher-diffraction-order components obtained in the process of Fresnel diffraction of a Gaussian beam by forked grating [22]. It differs by the transmission coefficients and by the polar coordinates $\rho$ and $\theta$ (instead of $\rho_{\pm m}$ and $\theta_{\pm m}$ present in [22]), since, here, all the diffracted components are coaxial.

There is another specific phenomenon in this case: the sum of the two wave fields, diffracted in the (+$m$)th and (-$m$)th diffraction order

$$U_{+m}(\rho,\theta,z) + U_{-m}(\rho,\theta,z) = (-i)^m J_m(k\delta)\exp(-ik\delta)\sqrt{\pi}\,\frac{w_0}{w(z)}\left\{\frac{1}{w^2(z)} - \frac{ik}{2}\left(\frac{1}{z} - \frac{1}{R(z)}\right)\right\}^{1/2}\exp[-i\phi(z)]$$

$$\times \cos[mp(\theta + \pi/2)]\rho\exp\left(\frac{-\rho^2}{2w^2(z)}\right)\left\{I_{(mp-1)/2}\left(\frac{x}{2}\right) - I_{(mp+1)/2}\left(\frac{x}{2}\right)\right\}, \tag{12}$$

shows an azimuthal modulation of the bright ring, which surrounds the central dark spot, in the transverse intensity profile. The first ring, azimuthally modulated, consists of $2p$ bright spots. The resulting wave field $U_{+m} + U_{-m}$ has a dark axis. But, it does not possess a phase term of type $\exp(ip\varphi)$, meaning that it is not a vortex beam.

Finally, the total diffracted wave field, according to Eq. (5), is represented as

$$U(\rho,\theta,z) = \frac{w_0}{w(z)}\exp(-ik\delta)\exp\left\{-ik\left[z + \frac{\rho^2}{2R(z)} - \frac{1}{k}\arctan(z/z_0)\right]\right\}$$

$$\times\left\{\exp\left(\frac{-\rho^2}{w^2(z)}\right)J_0(k\delta) + \sqrt{\pi}\left[\frac{-ik}{2}\left(\frac{1}{z}-\frac{1}{R(z)}\right) + \frac{1}{w^2(z)}\right]^{1/2}\exp\left[\frac{ik\rho^2}{4}\left(\frac{1}{R(z)}-\frac{1}{z}\right)\right]\right.$$

$$\left.\times\rho\exp\left(\frac{-\rho^2}{2w^2(z)}\right)\sum_{m=1}^{\infty}(-i)^m J_m(k\delta)\cos(mp(\theta+\pi/2))\left[I_{(mp-1)/2}\left(\frac{x}{2}\right)-I_{(mp+1)/2}\left(\frac{x}{2}\right)\right]\right\}, \quad (13)$$

with modified Bessel functions argument given by Eq. (10).

The intensity distribution, for the case of satisfying the condition (11), is calculated as

$$I(\rho,\theta,z) = |U(\rho,\theta,z)|^2 = \frac{w_0^2}{w^2(z)}\pi\left[\frac{1}{w^4(z)} + \frac{k^2}{4}\left(\frac{1}{z}-\frac{1}{R(z)}\right)^2\right]^{1/2}$$

$$\times\rho^2\exp\left(\frac{-\rho^2}{w^2(z)}\right)\left|\sum_{m=1}^{\infty}(-i)^m J_m(k\delta)\cos(mp(\theta+\pi/2))\left[I_{(mp-1)/2}\left(\frac{x}{2}\right)-I_{(mp+1)/2}\left(\frac{x}{2}\right)\right]\right|^2. \quad (14)$$

The "sheaf" of the coupled beams (14) dictates a slowly divergent distribution. This can also be noticed in Fig. 2, where the calculated transverse intensity profiles at different $z$-distances are shown: the central dark spot is wider at bigger $z$-distance. The numerical calculation is done based on Eq. (14), and using the following parameters of the cosine-profiled phase Siemens star: $p=9$, $n=1,48$ for $\lambda=1$ μm, $c=10$ μm ($k\delta=15$), while the beam waist radius is $w_0=3,5$ mm. The horizontal and vertical coordinates in Fig. 2 are given in mm. There is no angular rotation of the transverse intensity pattern as the beam propagates in $z$-direction.

In order to transform these beams into nondiverging ones, an axicon will be joined to the CPSS, and we'll refer to the new diffractive element as a cosine-profiled phase Siemens star axicon (CPSSA).

## 3. COSINE-PROFILED PHASE SIEMENS STAR AXICON UNDER GAUSSIAN LASER BEAM ILLUMINATION

Our further purpose is to investigate theoretically the hybrid optical element composed by a very thin axicon with a base radius $R_0$ and transmittance $T(r) = \exp(ik\alpha_0 r)$, and a cosine-profiled phase Siemens star with transmittance $T(\varphi) = \exp\{-ik[\delta + \delta\cos(p\varphi)]\}$. Thus, the transmission function of this hybrid optical element, named as cosine-profiled phase Siemens star axicon, in thin transparency approximation, can be defined as

$$T'(r,\varphi) = A(r)\exp(ik\alpha_0 r)\exp\{-ik[\delta + \delta\cos(p\varphi)]\}. \tag{15}$$

In Eq. (15) the parameter $\alpha_0 = (n'-1)\gamma$ is connected to the axicon base angle $\gamma$ (for which the approximation $\tan\gamma \approx \sin\gamma \approx \gamma$ is valid) and its refractive index $n'$, while the integer $p$ is the spatial frequency in azimuthal direction of the CPSS and of the CPSSA. The function $A(r)$ plays the role of a beam truncation function and is defined as

$$A(r) = \begin{cases} 1 & \text{when } w_0 < R_0, \\ \text{circ}(r/R_0) & \text{when } w_0 \geq R_0, \end{cases}$$

where $\text{circ}\left(\dfrac{r}{R_0}\right) = \begin{cases} 1, & \text{when } r \leq R_0 \\ 0, & \text{when } r > R_0. \end{cases}$ The cosine-profiled phase Siemens star is attached to the axicon base situated in the plane $\Delta(r,\varphi)$. Normally to this plane, a Gaussian beam $U^{(i)}(r,\varphi,0) = \exp(-r^2/w_0^2)$, whose waist is in this plane, is entering, and its propagation axis is passing through the center of the cosine-profiled phase Siemens star axicon (Fig. 3).

### *Diffraction of a Gaussian laser beam by CPSSA*

In order to obtain the analytical expression for the diffracted wave field $U(\rho,\theta,z)$ in the plane $\Pi(\rho,\theta)$, a distance $z$ from the diffractive optical element plane $\Delta(r,\varphi)$, we solve the Fresnel diffraction integral written in polar coordinates, in which the transmission function of the CPSSA, $T'(r,\varphi)$, and the incident beam expression $U^{(i)}(r,\varphi)$ are inserted. After integration over the azimuthal variable (done in Section 2), we get

$$U(\rho,\theta,z) = \frac{ik}{z}\exp[-ik(z + \rho^2/2z + \delta)]$$
$$\times \left\{ J_0(k\delta)Y_0(r) + \sum_{m=1}^{\infty}(-i)^m i^{mp} J_m(k\delta)Y_m(r)[\exp(+imp\theta) + \exp(-imp\theta)] \right\}, \tag{16}$$

with integrals over the radial coordinate now denoted as

$$Y_0(r) = \int_0^{\infty} A(r)\exp\left[-ik\left(\frac{r^2}{2z} - \alpha_0 r\right)\right] J_0\left(\frac{kr\rho}{z}\right)\exp\left(\frac{-r^2}{w_0^2}\right) r\,dr, \tag{17}$$

$$Y_m(r) = \int_0^{\infty} A(r)\exp\left[-ik\left(\frac{r^2}{2z} - \alpha_0 r\right)\right] J_{mp}\left(\frac{kr\rho}{z}\right)\exp\left(\frac{-r^2}{w_0^2}\right) r\,dr. \tag{18}$$

Their solutions are found by means of the stationary phase method [23, 24], and in both cases are evaluated around the stationary point $r_c = \alpha_0 z$, as

$$Y_0(r_c) = \alpha_0 z \sqrt{\frac{z}{k}} A(r_c) J_0(k\alpha_0 \rho) \exp\left[\frac{-z^2}{(w_0/\alpha_0)^2}\right] \exp\left(\frac{ikz\alpha_0^2}{2}\right), \tag{19}$$

$$Y_m(r_c) = \alpha_0 z \sqrt{\frac{z}{k}} A(r_c) J_{mp}(k\alpha_0 \rho) \exp\left[\frac{-z^2}{(w_0/\alpha_0)^2}\right] \exp\left(\frac{ikz\alpha_0^2}{2}\right). \tag{20}$$

By substituting these expressions in Eq. (16) the total diffracted wave field is written in the form

$$U(\rho,\theta,z) = i\sqrt{\frac{w_0 \alpha_0 k}{\sqrt{2}}} A(r_c) \left(\frac{z\sqrt{2}}{w_0/\alpha_0}\right)^{1/2} \exp\left(\frac{-z^2}{w_0^2/\alpha_0^2}\right) \exp\left[-ik\left(z - \frac{\alpha_0^2}{2}z + \frac{\rho^2}{2z} + \delta\right)\right]$$
$$\times \left\{ J_0(k\delta) J_0(k\alpha_0 \rho) + 2\sum_{m=1}^{\infty}(-1)^m i^{m(p+1)} J_m(k\delta) J_{mp}(k\alpha_0 \rho) \cos(mp\theta) \right\}. \tag{21}$$

If $p=2p'-1$ ($p'$ is an integer) is satisfied, the expression in the sum over the index $m$ in Eq. (21) becomes real, and Eq. (21) transforms into

$$U(\rho,\theta,z) = i\sqrt{\frac{w_0 \alpha_0 k}{\sqrt{2}}} A(r_c) \left(\frac{z\sqrt{2}}{w_0/\alpha_0}\right)^{1/2} \exp\left(\frac{-z^2}{w_0^2/\alpha_0^2}\right) \exp\left[-ik\left(z - \frac{\alpha_0^2}{2}z + \frac{\rho^2}{2z} + \delta\right)\right]$$
$$\times \left\{ J_0(k\delta) J_0(k\alpha_0 \rho) + 2\sum_{m=1}^{\infty}(-1)^{m(p'+1)} J_m(k\delta) J_{m(2p'-1)}(k\alpha_0 \rho) \cos(m(2p'-1)\theta) \right\}. \tag{22}$$

Now $A(r_c) = \begin{cases} 1 & \text{when } w_0 < R_0, \\ \text{circ}(r_c/R_0) = \text{circ}(\alpha_0 z/R_0) & \text{when } w_0 \geq R_0. \end{cases}$

The total diffracted wave field consists of coaxial wave components which are nondiverging (the argument of the Bessel functions $J_0(k\alpha_0 \rho)$ and $J_{m(2p'-1)}(k\alpha_0 \rho)$ does not depend on the $z$-coordinate).

The coaxial optical beams $U_{+m}$ and $U_{-m}$ in the higher diffraction orders are vortex ones and carry topological charges $+mp$ and $-mp$, respectively. But, as in Sec. 2, because of the coupling effect, their optical charges transform into an azimuthal-cosine characteristic of the common wave amplitude, which is not vortex.

The previous result can be simplified by choosing the value of $k\delta$ to coincide with one of the roots of the Bessel function $J_0(k\delta)$ (meaning that the bright-axis, zeroth-diffraction-order beam is eliminated). In that case, according to Eq. (22), the intensity distribution will be given as

$$I(\rho,\theta,z) = \frac{4w_0\alpha_0 k}{\sqrt{2}} A(r_c) \left(\frac{2z^2}{(w_0/\alpha_0)^2}\right)^{1/2} \exp\left(\frac{-2z^2}{(w_0/\alpha_0)^2}\right)$$
$$\times \left[\sum_{m=1}^{\infty} (-1)^{m(p'+1)} J_m(k\delta) J_{m(2p'-1)}(k\alpha_0\rho) \cos(m(2p'-1)\theta)\right]^2, \quad (23)$$

which we also write in the form

$$I(\rho,\theta,z) = \frac{4w_0\alpha_0 k}{\sqrt{2}} A(r_c) I^{1/2}(z, w_0/\alpha_0)$$
$$\times \left[\sum_{m=1}^{\infty} (-1)^{m(p'+1)} J_m(k\delta) J_{m(2p'-1)}(k\alpha_0\rho) \cos(m(2p'-1)\theta)\right]^2. \quad (24)$$

In the transverse intensity profile, the central dark spot will be surrounded by a bright, ruptured ring (due to the azimuthal variation of the intensity, which has $2p$ zeroes for a cycle from 0 to $2\pi$). It means that a bright spots array around the dark, nonvortex central core will occur.

In Fig. 4, the transverse intensity profiles of the diffracted wave field, computer-generated based on Eq. (23), at distance $z$=16 cm from the diffractive optical element, and for two different spatial frequencies, $p$=3 (a) and $p$=9 (b), are shown. The parameters for the cosine-profiled phase Siemens star axicon used are: $\gamma$=1,35$^0$=0,0235 rad, $n=n'$=1,48 for incident beam wavelength $\lambda$=1 μm, $c$=10 μm ($k\delta$=15), and the beam waist radius is $w_0$=3,5 mm. The horizontal and vertical coordinates in Fig. 4 (as well as in the forthcoming Fig. 6) are given in mm. The resulting wave amplitude $U$ has a dark axis, but it does not possess a phase singularity. The central dark spot is surrounded by $2p$ bright spots, situated in a circular array. One can see that the central, nondiverging dark spot has a bigger radius for bigger value of $p$. The dark spaces in this array are narrower than the bright spaces and they are suitable for angular alignment. The transverse intensity profiles do not rotate with change of the $z$-distance.

The nondiverging beams, obtained in the zeroth and in the higher diffraction orders, are coaxial and have same propagation distances, which are defined as in [24]:

a) $L_{max} = R_0/\alpha_0$, for the case when $w_0 \geq R_0$; then, the function $A(r_c) = \text{circ}\left(\dfrac{z}{(R_0/\alpha_0)}\right)$ dictates the longitudinal truncation of the outgoing beam.

b) $L_{max} = \dfrac{w_0}{\alpha_0}\sqrt{\dfrac{3}{2}}$, when $w_0 < R_0$. In this case the propagation distance is defined by the function $I^{1/2}(z, w_0/\alpha_0) = \left(\dfrac{2z^2}{(w_0/\alpha_0)^2}\right)^{1/2} \exp\left(\dfrac{-2z^2}{(w_0/\alpha_0)^2}\right)$ in Eq. (24), which is of Gauss-doughnut type, similar to the $\left(\dfrac{2r^2}{w_0^2}\right)^l \exp\left(\dfrac{-2r^2}{w_0^2}\right)$-radial irradiance distribution of the Laguerre-Gaussian $LG_0^l$ beam (with zeroth radial mode number and azimuthal mode number $l$). But, instead to the radial coordinate $r$, it is related to the axial coordinate $z$, and, also there is a difference in its definition by the half integer power. Similarly, as the radial irradiance function of the beam defines the outside limit of its ring (the beam transverse profile size) as $\sigma_0^l = w_0\sqrt{l+1}$ [25], the axial intensity distribution determines the beam attenuation distance as $L_{max} = \dfrac{w_0}{\alpha_0}\sqrt{l+1+1/2} = \dfrac{w_0}{\alpha_0}\sqrt{3/2}$. The maximum intensity in the brightest places in the spots along the first circular array (at angular positions where $\cos((2p'-1)\theta) = 1$ is satisfied) is changing along the $z$-propagation distance as it is shown in Fig. 5 (where with $I_{norm}$ the normalized intensity is denoted). According to Eq. (32) in [24], the bright spots have maximum intensity at distance $z_{max} = w_0/2\alpha_0$ (in this case we replace $l=0$ since the incident beam is Gaussian).

The intensity in the first array of bright spots, at given $z$-distance, can be approximated as proportional to $J_{2p'-1}^2(k\alpha_0\rho)\cos^2((2p'-1)\theta)$, because the higher-order Bessel beams have their first bright rings more distant from the center, where the interference effect between them is predominant. Therefore, the radius of this ruptured ring surrounding the dark core can be evaluated as: $\rho = \mu_{p',1}/k\alpha_0$, with $\mu_{p',1}$ being the value of the argument of the Bessel function $J_{2p'-1}^2(k\alpha_0\rho)$, for which the first derivative defines its first maximum. It can be calculated as a root of the equation: $J_{2p'}(k\alpha_0\rho) = J_{2(p'-1)}(k\alpha_0\rho)$.

In Fig. 6 the transverse intensity diffraction pattern for the case of presence of the zeroth-diffraction-order is shown. The computer calculation is done by using the expression for the diffracted wave field $U(\rho,\theta,z)$ given by Eq. (22), and taking into consideration that the intensity is: $I(\rho,\theta,z) \propto |U(\rho,\theta,z)|^2$, for $z$=16 cm. The rest of the parameters values are the same as those used for obtaining Fig. 4. b, except now $k\delta$=13. In this case, the radius of the central bright spot can be calculated as: $\rho = 2,4/k\alpha_0$, where 2,4 is the first zero of the function $J_0^2(k\alpha_0\rho)$.

## 4. CONCLUSIONS

We have investigated the Fresnel diffraction of a Gaussian laser beam by a phase cosine-profiled Siemens star, entering with its waist in this optical element plane. The output field is found as a sum of coaxial zeroth-diffraction-order, chargeless Gaussian beam, and higher-diffraction-order beams, possessing topological charges $\pm mp$ in the $(\pm m)$th diffraction orders (where $p$ is the spatial frequency of the CPSS). But, as a result of the interference between the opposite charged positive and negative $m$th diffraction orders, the total diffracted wave field does not possess a phase singularity. When the zeroth-diffraction-order beam is eliminated, the transverse intensity profile of the output beam has a central dark spot surrounded by bright spots, arranged along circle with double spatial frequency, which diverges.

Further, we propose as an optical element a combination of a CPSS and an axicon, which we name as cosine-profiled phase Siemens star axicon. The theoretical investigation of the problem of Fresnel diffraction of a Gaussian laser beam passing through the center of the CPSSA with its waist in this optical element plane, shows that the output field is similar to that obtained in the case of CPSS, but, now it preserves its transverse intensity profile with same dimensions along defined propagation interval. Because of this, the intensity distribution of the nondiverging "sheaf" of coupled beams (Eq. (22)), as it is the case with the specially designed masks in [17], or the binary Siemens star [14], treated as an optical diffractive element in [10], can be used for high-precision alignment in some optical and mechanical engineering.

**REFERENCES**

[1] J. Dyson, "Circular and Spiral Diffraction Gratings," Proceedings of the Royal Society of London, Series A, Mathematical and Physical Sciences **248**, 93-106 (1958).

[2] J. Durnin, "Exact solutions for non-diffracting beams. I. The scalar theory," J. Opt. Soc. Am. A **4**, 651–654 (1987).

[3] J. Durnin, J. J. Miceli, Jr., and J. H. Eberly, "Diffraction-free beams," Phys. Rev. Lett. **58**, 1499-1501 (1987).

[4] F. Gori, G. Guattari, and C. Padovani, "Bessel-Gauss beams," Opt. Commun. **64**, 491-495 (1987).

[5] D. McGloin and K. Dholakia, "Bessel beams: diffraction in a new light," Contemporary Physics **46**, 15 – 28 (2005).

[6] J. H. McLeod, "The axicon: a new type of optical element," J. Opt. Soc. Am. **44**, 592–597 (1954).

[7] R. M. Herman and T. A. Wiggins, "Production and uses of diffractionless beams," J. Opt. Soc. Am. A **8**, 932-942 (1991).

[8] J. Turunen, A. Vasara, and A. T. Friberg, "Holographic generation of diffraction-free beams," Appl. Opt. **27**, 3959-3962 (1988).

[9] A. C. S. van Heel, "Modern alignment devices," in *Progress in Optics*, *Vol. 1* (Springer-Verlag, New York, 1961), pp. 289-329.

[10] J. Ojeda-Castañeda, P. Andrés, and M. Martínez-Corral, "Zero axial irradiance by annular screens with angular variation," Appl. Opt. **31**, 4600-4602 (1992).

[11] V. V. Kotlyar, A. A. Almazov, S. N. Khonina, V. A. Sofier, H. Elfstrom, and J. Turunen, "Generation of phase singularity through diffracting a plane or Gaussian beam by a spiral phase plate", J. Opt. Soc. Am. A **22**, 849-861 (2005).

[12] M. W. Beijersbergen, R. P. C. Coerwinkel, M. Kristensen, and J. P. Woerdman, "Helical-wavefront laser beams produced with a spiral phase plate", Opt. Commun. **112**, 321-327 (1994).

[13] S. N. Khonina, V. V. Kotlyar, V. A. Soifer, K. Jefimovs, and J. Turunen, "Generation and selection of laser beams represented by a superposition of two angular harmonics," J. Mod. Opt. **51**, 761-773 (2004).

[14] E. B. Brown, *Modern Optics* (Reinhold Publishing Corporation, New York, 1965).

[15] C. Holzner, M. Feser, S. Vogt, B. Hornberger, S. B. Baines, and C. Jacobsen, "Zernike phase contrast in scanning microscopy with X-rays," Nature Physics **6**, 883–887 (2010).

[16] G. Kopitkovas, T. Lippert, C. David, A. Wokaun, and J. Gobrecht, "Fabrication of micro-optical elements in quartz by laser induced backside wet etching," Microelectronic Engin. **67-68**, 438-444 (2003).

[17] J. A. Davis, E. Carcole, and D. M. Cottrell, "Nondiffracting interference patterns generated with programmable spatial light modulators," Appl. Opt. **35**, 599-602 (1996).

[18] S. N. Khonina, V.V. Kotlyar, V.A. Soifer, M.V. Shinkaryev, and G.V. Uspleniev, "Trochoson," Opt. Commun. **91**, 158-162 (1992).

[19] M. Abramowitz and I. A. Stegun, *Handbook of Mathematical Functions* (Dover publ. Inc., New York, 1964).

[20] M. Born and E. Wolf, *Principles of Optics* (Cambridge, University Press, 1999).

[21] A. P. Prudnikov, Y. A. Brichkov, and O. I. Marichev, *Integrals and Series; Special Functions* (Nauka, Moskva, 1983).

[22] Lj. Janicijevic and S. Topuzoski, "Fresnel and Fraunhofer diffraction of a Gaussian laser beam by fork-shaped gratings," J. Opt. Soc. Am. A **25**, 2659-2669 (2008).

[23] J. A. Davis, E. Carcole, and D. M. Cottrell, "Intensity and phase measurements of nondiffracting beams generated with a magneto-optic spatial light modulator," Appl. Opt. **35**, 593-598 (1996).

[24] S. Topuzoski and Lj. Janicijevic, "Conversion of high order Laguerre-Gaussian beams into Bessel beams of increased, reduced or zero-th order by use of a helical axicon," Opt. Commun. **282**, 3426-3432 (2009).

[25] R. L. Phillips and L. C. Andrews, "Spot size and divergence for Laguerre-Gaussian beams of any order," Appl. Opt. **22**, 643-644 (1983).

**Figure captions:**

Fig. 1. Cosine change of the phase transmission function of the CPSS in azimuthal direction. The spatial frequency $p=11$.

a) Two-dimensional representation in polar coordinates.

b) One-dimensional representation of the phase variation due to the cosine change of the thickness of the CPSS.

Fig. 2. Transverse intensity distribution calculated on base on Eq. (14) at different distances: $z=60$ mm (a), $z=100$ mm (b), $z=160$ mm (c).

Parameters used for the cosine-profiled phase Siemens star are: $p=9$, $n=1,48$ for $\lambda=1$ µm, $c=10$ µm ($k\delta = 15$). The beam waist radius is $w_0=3,5$ mm. The zoomed central part of Fig. 2. a is shown in Fig. 2. a'.

Fig. 3. The geometry of the problem.

Fig. 4: Transverse intensity profile of the diffracted beam by the CPSSA, according to Eq. (23), for: $p=3$ (a) and $p=9$ (b).

Fig. 5. Intensity variation on the bright spots situated in a circle around the central dark spot, along the beam propagation direction $z$, according to Eq. (23).

Fig. 6. Transverse intensity profile of the diffracted beam (22) by the CPSSA when the zeroth-diffraction-order beam is not eliminated. The parameters used are: $p=9$, $z=16$ cm, $\gamma=0,0235$ rad, $n=n'=1,48$ for incident beam wavelength $\lambda=1$ µm, $k\delta=13$, $w_0=3,5$ mm.

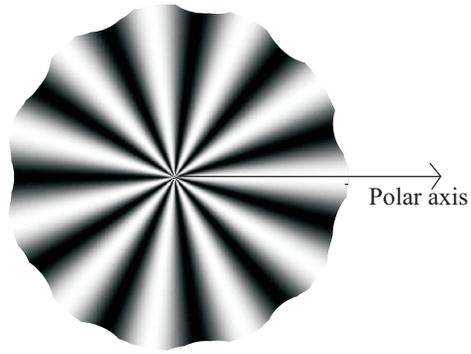

Fig. 1. a.

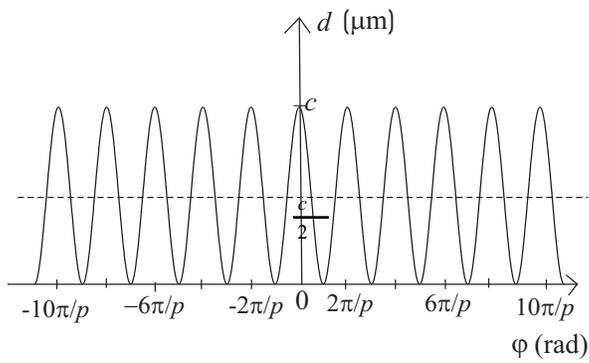

Fig. 1. b.

Fig. 1. Cosine change of the phase transmission function of the cosine-profiled phase Siemens star (CPSS) in azimuthal direction. The spatial frequency $p$=11.

a) Two-dimensional representation in polar coordinates.

b) One-dimensional representation of the phase variation due to the cosine change of the thickness of the CPSS.

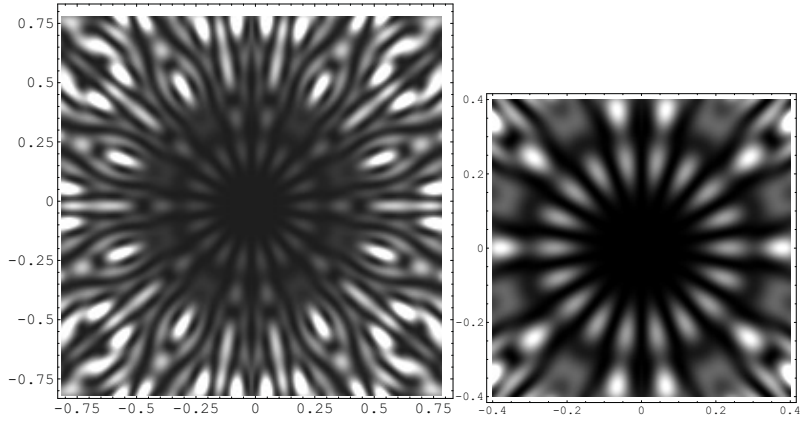

Fig. 2. a.                    Fig. 2. a'.

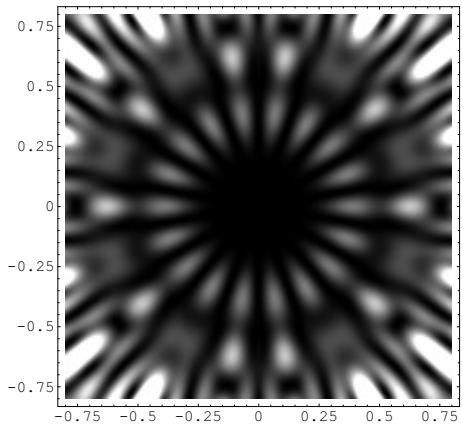

Fig. 2. b.

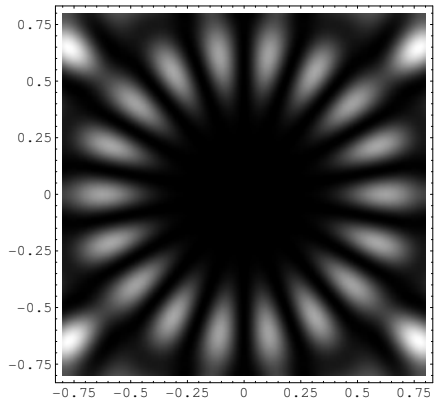

Fig. 2. c.

Fig. 2. Transverse intensity distribution calculated on base on Eq. (14) at different distances: $z$=60 mm (a), $z$=100 mm (b), $z$=160 mm (c). Parameters used for the cosine-profiled phase Siemens star

are: $p=9$, $n=1,48$ for $\lambda=1$ μm, and $c=10$ μm ($k\delta=15$). The beam waist radius is $w_0=3,5$ mm. The zoomed central part of Fig. 2. a is shown in Fig. 2. a'.

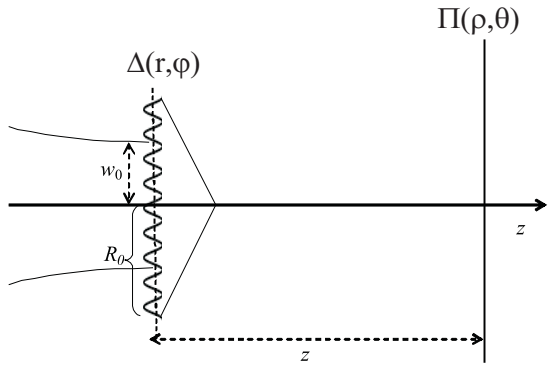

Fig. 3. The geometry of the problem.

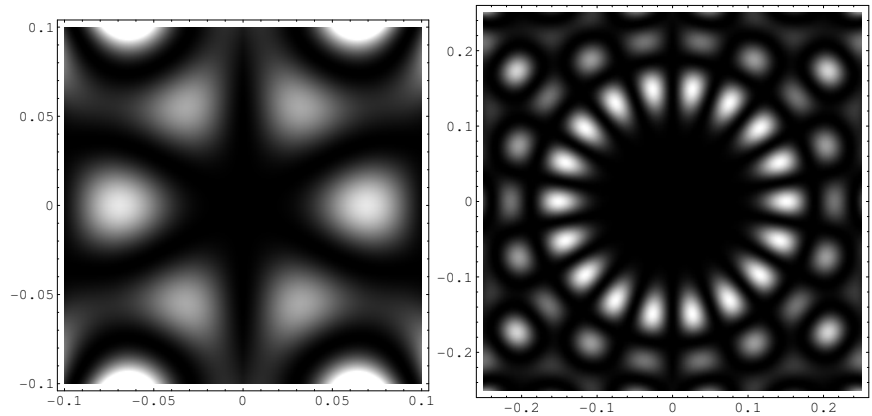

Fig. 4. a            Fig. 4. b.

Fig. 4: Transverse intensity profile of the diffracted beam by the CPSSA, according to Eq. (23), for: $p=3$ (a) and $p=9$ (b).

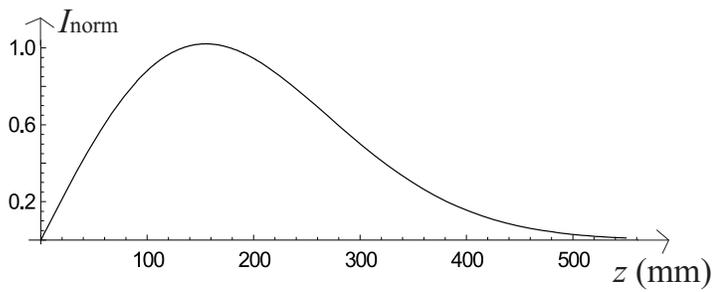

Fig. 5. Intensity variation on the bright spots situated in a circle around the central dark spot, along the beam propagation direction $z$, according to Eq. (23).

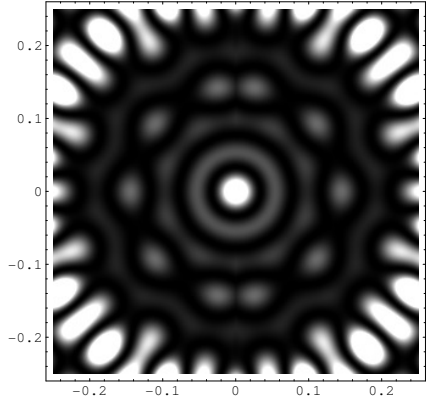

Fig. 6. Transverse intensity profile of the diffracted beam (22) by the CPSSA when the zeroth-diffraction-order beam is not eliminated. The parameters used are: $p=9$, $z=16$ cm, $\gamma=0,0235$ rad, $n=n'=1,48$ for incident beam wavelength $\lambda=1$ μm, $k\delta=13$, and $w_0=3,5$ mm.